# A spin operator for modeling ferromagnetic alignment.

Akande R. O*, Oyewande O.E

*Theoretical Physics Group,*

*Department of Physics, University of Ibadan, Ibadan, Nigeria.*

*telleverybodythat@gmail.com*

**Abstract**

We have limited our interaction scope to those involving particle's spin. We then develop operators for handling a many-body treatment of the spin interactions. The results presented show how to promote the particles of the system into the same spin state with time. We therefore, in order to tune the model for ferromagnetism, assume that the operators act only on particles that are, initially, of the same spin state. The observed growth in the population of particles of the same spin with time makes the model a success. Almost all of the recent experimentally observed and theoretically predicted formations are accurately obtainable with this model.

# 1. Introduction

We start by representing a symmetric spin state of a system of particles with a matrix, equation (1). The system is believed to be in the same spin state such that all the particles of the system are either in up or down state. Such system will have, at any point in time, the same number of particles in up and down states.

$$S_z = \begin{pmatrix} 1 & 0 & 1 & 0 \\ 0 & 1 & 0 & 1 \\ 1 & 0 & 1 & 0 \\ 0 & 1 & 0 & 1 \end{pmatrix} \quad \text{...............(1)}$$

In the equation (1) above, we have represented spin up with "1" and down with "0" [4] [5], where the orientation along the z-axis is taken as spin up. The equation represents what we call the default ground spin state of the particles. It is bound to happen when the particles are either in very low temperature or close together in an interaction condition. This is the spin state they always want to and eventually return in ultra cold scenario. However, temperatures of everyday live prevent this and the order or symmetry is, hence, lowered or broken. There could be many ways to handle the very cumbersome interaction mathematically but the easiest is and has always been application of matrices [1]. If one observes carefully, we could see the spread of cosine (or sine) waveforms along every single straight line in the ensemble.

If we change the orientation of any one of the particle, the resultant spin just changes its direction opposite to the former and nothing more. Whatever the particle chosen and whatever its new spin orientation is, will only change the direction of net spin from +cosine waveform to –cosine waveform and vice versa [2]. The only time where there will be a no net spin waveform (but net spin direction) will be if any of the particles have relatively larger Coulomb field (or separation distance) such that it prevents them from interacting via spin. This will be the source of the action of one our operators. Imagine particles 2 and 6 have a relatively larger Coulomb, at a moment of time, than others, both are spaced out and thus do not interact via spin. An application of this knowledge is that there can only be net spin direction (up or down) if and only if the particles have a gradual and sequential absorption of external energy. Otherwise, the net spin may cancel out.

# 2. Methodology

Stating the operator to be [3] : $\tau_{ij}^{\dagger} = \sum_{n=1}^{i,j} \left( a_{i,j\pm n} \oplus a_{i\pm n,j} \right)$ ........( 2 ) ,we shall strongly restrict the operation to be in the forward direction (and not backwards) and that it changes the spin of the nearest neighbors of a selected particle $a_{i,j}$. We can easily show that the operator $\tau_{ij}^{\dagger}$ picks particles along the vertical and horizontal which are given by $a_{i,j\pm n}$ and $a_{i\pm n,j}$ respectively. It is very important to make sure that the number of spin ups (1) and downs (0) are the same and for the purposes of near-reality many-body treatment, we will consider a large matrix such as a 10 x 10 matrix.

$$\vec{S}_z = \begin{pmatrix} 1 & 0 & 1 & 0 & 1 & 0 & 1 & 0 & 1 & 0 \\ 0 & 1 & 0 & 1 & 0 & 1 & 0 & 1 & 0 & 1 \\ 1 & 0 & 1 & 0 & 1 & 0 & 1 & 0 & 1 & 0 \\ 0 & 1 & 0 & 1 & 0 & 1 & 0 & 1 & 0 & 1 \\ 1 & 0 & 1 & 0 & 1 & 0 & 1 & 0 & 1 & 0 \\ 0 & 1 & 0 & 1 & 0 & 1 & 0 & 1 & 0 & 1 \\ 1 & 0 & 1 & 0 & 1 & 0 & 1 & 0 & 1 & 0 \\ 0 & 1 & 0 & 1 & 0 & 1 & 0 & 1 & 0 & 1 \\ 1 & 0 & 1 & 0 & 1 & 0 & 1 & 0 & 1 & 0 \\ 0 & 1 & 0 & 1 & 0 & 1 & 0 & 1 & 0 & 1 \end{pmatrix} \quad \ldots\ldots(3)$$

For instance, let us pick particle $a_{i,j}$ and write the operation of (2) on matrix (3) as

$$\bar{\vec{S}}_z = \tau_{ij}^\dagger \vec{S}_z \ldots\ldots\ldots(4) \text{ [1]}$$

The new spin waveform, as shown below, will spread out from the particle chosen and the entire new spin waveform symmetry looks much like the inverted form of the first. The new spin symmetry will still be unbroken, but just inverted.

$$\bar{\vec{S}}_z = \tau_{65}^\dagger \vec{S}_z = \begin{pmatrix} 0 & 1 & 0 & 1 & 0 & 1 & 0 & 1 & 0 & 1 \\ 1 & 0 & 1 & 0 & 1 & 0 & 1 & 0 & 1 & 0 \\ 0 & 1 & 0 & 1 & 0 & 1 & 0 & 1 & 0 & 1 \\ 1 & 0 & 1 & 0 & 1 & 0 & 1 & 0 & 1 & 0 \\ 0 & 1 & 0 & 1 & 0 & 1 & 0 & 1 & 0 & 1 \\ 1 & 0 & 1 & 0 & 1 & 0 & 1 & 0 & 1 & 0 \\ 0 & 1 & 0 & 1 & 0 & 1 & 0 & 1 & 0 & 1 \\ 1 & 0 & 1 & 0 & 1 & 0 & 1 & 0 & 1 & 0 \\ 0 & 1 & 0 & 1 & 0 & 1 & 0 & 1 & 0 & 1 \\ 1 & 0 & 1 & 0 & 1 & 0 & 1 & 0 & 1 & 0 \end{pmatrix} \quad \ldots\ldots(5)$$

We can have many of such operators working in turn. An example is presented below:

$$\tau_{11}^{\dagger}\tau_{87}^{\dagger};\vec{S}_z = \begin{pmatrix} 0 & 1 & 0 & 1 & 0 & 1 & 0 & 1 & 0 & 1 \\ 1 & 0 & 1 & 0 & 1 & 0 & 1 & 0 & 1 & 0 \\ 0 & 1 & 0 & 1 & 0 & 1 & 0 & 1 & 0 & 1 \\ 1 & 0 & 1 & 0 & 1 & 0 & 1 & 0 & 1 & 0 \\ 0 & 1 & 0 & 1 & 0 & 1 & 0 & 1 & 0 & 1 \\ 1 & 0 & 1 & 0 & 1 & 0 & 1 & 0 & 1 & 0 \\ 0 & 1 & 0 & 1 & 0 & 1 & 0 & 1 & 0 & 1 \\ 1 & 0 & 1 & 0 & 1 & 0 & 1 & 0 & 1 & 0 \\ 0 & 1 & 0 & 1 & 0 & 1 & 0 & 1 & 0 & 1 \\ 1 & 0 & 1 & 0 & 1 & 0 & 1 & 0 & 1 & 0 \end{pmatrix} = \bar{\vec{S}}_z \quad \text{........(6)}$$

It is important to state that the semi colon in the last equation is to lay emphasis on the order in which the operation is performed.

$$\tau_{11}^{\dagger};(\tau_{87}^{\dagger}\vec{S}_z) = \begin{pmatrix} 1 & 0 & 1 & 0 & 1 & 0 & 1 & 0 & 1 & 0 \\ 0 & 1 & 0 & 1 & 0 & 1 & 0 & 1 & 0 & 1 \\ 1 & 0 & 1 & 0 & 1 & 0 & 1 & 0 & 1 & 0 \\ 0 & 1 & 0 & 1 & 0 & 1 & 0 & 1 & 0 & 1 \\ 1 & 0 & 1 & 0 & 1 & 0 & 1 & 0 & 1 & 0 \\ 0 & 1 & 0 & 1 & 0 & 1 & 0 & 1 & 0 & 1 \\ 1 & 0 & 1 & 0 & 1 & 0 & 1 & 0 & 1 & 0 \\ 0 & 1 & 0 & 1 & 0 & 1 & 0 & 1 & 0 & 1 \\ 1 & 0 & 1 & 0 & 1 & 0 & 1 & 0 & 1 & 0 \\ 0 & 1 & 0 & 1 & 0 & 1 & 0 & 1 & 0 & 1 \end{pmatrix} = \vec{S}_z \quad \text{........(7)}$$

As we have seen, from equations (6-7), whatever the number of times the operation $\tau_{ij}^{\dagger}$ is performed on our matrix of symmetric spin system, the new symmetry is always the inverted form of the last. In otherwords, the operation has not increased the number of particles with the same spin. To do this we introduce another operator that works opposite the former one. This new operator denoted by $\tau_{ij}$ works such that the chosen particle does not change in spin and neither does it change the spin of its neighbors. Unlike equation (4) we cannot have $\bar{\vec{S}}_z = \tau_{ij}\vec{S}_z$, since operator $\tau_{ij}$ only retains the spin of the chosen particle. We can suggest that the operator $\tau_{ij}$ actually acts like the Newton's law for spin motion at any point in time, meaning that the particle in question remains in its current spin state until when acted upon by operator $\tau_{ij}^{\dagger}$. We shall start combining the operation of both $\tau_{ij}^{\dagger}$ and $\tau_{ij}$ on our symmetric spin system, but note that operations such as this $\tau_{ij}\tau_{ij}\vec{S}_z$ will not make any sense if i=j. Also, it is very important to impose a relationship between the sites i and j with these two conditions that:

$|i+j| = k'$ ...(8a) and $|i-j| = k$ ....(8b) so that our new expression becomes :

$$\bar{S}_z = \tau_{ij}^\dagger \tau_{jk} \vec{S}_z.$$

For many combinations, we have :

$$\bar{S}_z = \tau_{ab}^\dagger ... \tau_{cd}^\dagger \tau_{ef}^\dagger \tau_{fi} \tau_{dj} ... \tau_{bk} \vec{S}_z ......( 9 )$$

Where $i' = |e+f|, i = |e-f|$

$j' = |c+d|, j = |c-d|$ and $k' = |a+b|, k = |a-b|$

As an example, let us consider $\bar{S}_z = \tau_{32}^\dagger \tau_{25} \vec{S}_z$, we have

$$\tau_{32}^\dagger \tau_{25} \vec{S}_z = \begin{pmatrix} 0 & 1 & 0 & 1 & 0 & 1 & 0 & 1 & 0 & 1 \\ 1 & 0 & 1 & 0 & 0 & 0 & 1 & 0 & 1 & 0 \\ 0 & 1 & 0 & 1 & 0 & 1 & 0 & 1 & 0 & 1 \\ 1 & 0 & 1 & 0 & 1 & 0 & 1 & 0 & 1 & 0 \\ 0 & 1 & 0 & 1 & 0 & 1 & 0 & 1 & 0 & 1 \\ 1 & 0 & 1 & 0 & 1 & 0 & 1 & 0 & 1 & 0 \\ 0 & 1 & 0 & 1 & 0 & 1 & 0 & 1 & 0 & 1 \\ 1 & 0 & 1 & 0 & 1 & 0 & 1 & 0 & 1 & 0 \\ 0 & 1 & 0 & 1 & 0 & 1 & 0 & 1 & 0 & 1 \\ 1 & 0 & 1 & 0 & 1 & 0 & 1 & 0 & 1 & 0 \end{pmatrix} ......(10)$$

$$\tau_{32}^\dagger \tau_{21} \vec{S}_z = \begin{pmatrix} 0 & 1 & 0 & 1 & 0 & 1 & 0 & 1 & 0 & 1 \\ 0 & 0 & 1 & 0 & 1 & 0 & 1 & 0 & 1 & 0 \\ 0 & 1 & 0 & 1 & 0 & 1 & 0 & 1 & 0 & 1 \\ 1 & 0 & 1 & 0 & 1 & 0 & 1 & 0 & 1 & 0 \\ 0 & 1 & 0 & 1 & 0 & 1 & 0 & 1 & 0 & 1 \\ 1 & 0 & 1 & 0 & 1 & 0 & 1 & 0 & 1 & 0 \\ 0 & 1 & 0 & 1 & 0 & 1 & 0 & 1 & 0 & 1 \\ 1 & 0 & 1 & 0 & 1 & 0 & 1 & 0 & 1 & 0 \\ 0 & 1 & 0 & 1 & 0 & 1 & 0 & 1 & 0 & 1 \\ 1 & 0 & 1 & 0 & 1 & 0 & 1 & 0 & 1 & 0 \end{pmatrix} .........(11)$$

From the last equations, we can see that the first symmetry breaking has just occurred and that the number of spin downs (because the particles at sites 25 and 21 are both at spin down states) has increased by one. For the increase in the number of particles with the same spin we consider larger spin operator combinations, since the action of operator $\tau_{ij}$ is actually responsible for the increment in the number of same spin particles. Now, we consider $\bar{S}_z = \tau^{\dagger}_{ab}\tau^{\dagger}_{cd}\tau_{di}\tau_{bj}\vec{S}_z$, [1]. As an example, we take care not to pick values that exceed the limit of the matrix, but if such happens we can neglect any higher values than 10 x 10 matrix limit. For instance, $\bar{S}_z = \tau^{\dagger}_{51}\tau^{\dagger}_{78}\tau_{81}\tau_{14}\vec{S}_z$ does not exist for the case $k' = |a+b|$ but exists for $k = |a-b|$.

$\bar{S}_z = \tau^{\dagger}_{51}\tau^{\dagger}_{43}\tau_{31}\tau_{14}\vec{S}_z$ and $\bar{S}_z = \tau^{\dagger}_{51}\tau^{\dagger}_{43}\tau_{37}\tau_{16}\vec{S}_z$ would give:

$$\tau^{\dagger}_{51}\tau^{\dagger}_{43}\tau_{31}\tau_{14}\vec{S}_z = \tau^{\dagger}_{51}(\tau^{\dagger}_{43};\tau_{31})\tau_{14}\vec{S}_z = \begin{pmatrix} 0 & 1 & 0 & 1 & 0 & 1 & 0 & 1 & 0 & 1 \\ 1 & 0 & 1 & 0 & 1 & 0 & 1 & 0 & 1 & 0 \\ 1 & 1 & 0 & 1 & 0 & 1 & 0 & 1 & 0 & 1 \\ 1 & 0 & 1 & 0 & 1 & 0 & 1 & 0 & 1 & 0 \\ 0 & 1 & 0 & 1 & 0 & 1 & 0 & 1 & 0 & 1 \\ 1 & 0 & 1 & 0 & 1 & 0 & 1 & 0 & 1 & 0 \\ 0 & 1 & 0 & 1 & 0 & 1 & 0 & 1 & 0 & 1 \\ 1 & 0 & 1 & 0 & 1 & 0 & 1 & 0 & 1 & 0 \\ 0 & 1 & 0 & 1 & 0 & 1 & 0 & 1 & 0 & 1 \\ 1 & 0 & 1 & 0 & 1 & 0 & 1 & 0 & 1 & 0 \end{pmatrix} \quad .....(12)$$

$$\tau^{\dagger}_{51}\tau^{\dagger}_{43}\tau_{31}\tau_{14}\vec{S}_z = \tau^{\dagger}_{51}(\tau^{\dagger}_{43}\tau_{31});\tau_{14}\vec{S}_z = \begin{pmatrix} 1 & 0 & 1 & 1 & 1 & 0 & 1 & 0 & 1 & 0 \\ 0 & 1 & 0 & 1 & 0 & 1 & 0 & 1 & 0 & 1 \\ 0 & 0 & 1 & 0 & 1 & 0 & 1 & 0 & 1 & 0 \\ 0 & 1 & 0 & 1 & 0 & 1 & 0 & 1 & 0 & 1 \\ 1 & 0 & 1 & 0 & 1 & 0 & 1 & 0 & 1 & 0 \\ 0 & 1 & 0 & 1 & 0 & 1 & 0 & 1 & 0 & 1 \\ 1 & 0 & 1 & 0 & 1 & 0 & 1 & 0 & 1 & 0 \\ 0 & 1 & 0 & 1 & 0 & 1 & 0 & 1 & 0 & 1 \\ 1 & 0 & 1 & 0 & 1 & 0 & 1 & 0 & 1 & 0 \\ 0 & 1 & 0 & 1 & 0 & 1 & 0 & 1 & 0 & 1 \end{pmatrix} \quad .....(13)$$

$$\tau_{51}^{\dagger}\tau_{43}^{\dagger}\tau_{37}\tau_{16}\vec{S}_z = \tau_{51}^{\dagger}(\tau_{43}^{\dagger}\tau_{37});\tau_{16}\vec{S}_z = \begin{pmatrix} 1 & 0 & 1 & 0 & 1 & 1 & 1 & 0 & 1 & 0 \\ 0 & 1 & 0 & 1 & 0 & 1 & 0 & 1 & 0 & 1 \\ 1 & 0 & 1 & 0 & 1 & 0 & 0 & 0 & 1 & 0 \\ 0 & 1 & 0 & 1 & 0 & 1 & 0 & 1 & 0 & 1 \\ 1 & 0 & 1 & 0 & 1 & 0 & 1 & 0 & 1 & 0 \\ 0 & 1 & 0 & 1 & 0 & 1 & 0 & 1 & 0 & 1 \\ 1 & 0 & 1 & 0 & 1 & 0 & 1 & 0 & 1 & 0 \\ 0 & 1 & 0 & 1 & 0 & 1 & 0 & 1 & 0 & 1 \\ 1 & 0 & 1 & 0 & 1 & 0 & 1 & 0 & 1 & 0 \\ 0 & 1 & 0 & 1 & 0 & 1 & 0 & 1 & 0 & 1 \end{pmatrix} \quad\ldots\ldots(14)$$

No increment of more particles into same spin states has been done. However, the lesson learnt is that for promotion of particles into same-spin state to be possible, the selected particles, upon which the operators act must be of the same spin initially. Otherwise, no promotion will be made. Another lesson learnt is that the combined operations like in equation (9) can be simplified by just allowing operator $\tau_{ij}$ to act on the already inverted symmetry made by $\tau_{ij}^{\dagger}$ and repeat the process for the next group of operation $\tau_{ij}^{\dagger}\tau_{ij}$.

Now, let us consider a case where the particles are initially of the same spin:

$$\tau_{13}^{\dagger}\tau_{51}^{\dagger}\tau_{14}\tau_{32}\vec{S}_z = \tau_{13}^{\dagger}(\tau_{51}^{\dagger}\tau_{14});\tau_{32}\vec{S}_z = \begin{pmatrix} 1 & 0 & 1 & 1 & 1 & 0 & 1 & 0 & 1 & 0 \\ 0 & 1 & 0 & 1 & 0 & 1 & 0 & 1 & 0 & 1 \\ 0 & 1 & 1 & 0 & 1 & 0 & 1 & 0 & 1 & 0 \\ 0 & 1 & 0 & 1 & 0 & 1 & 0 & 1 & 0 & 1 \\ 1 & 0 & 1 & 0 & 1 & 0 & 1 & 0 & 1 & 0 \\ 0 & 1 & 0 & 1 & 0 & 1 & 0 & 1 & 0 & 1 \\ 1 & 0 & 1 & 0 & 1 & 0 & 1 & 0 & 1 & 0 \\ 0 & 1 & 0 & 1 & 0 & 1 & 0 & 1 & 0 & 1 \\ 1 & 0 & 1 & 0 & 1 & 0 & 1 & 0 & 1 & 0 \\ 0 & 1 & 0 & 1 & 0 & 1 & 0 & 1 & 0 & 1 \end{pmatrix} \quad\ldots\ldots(15)$$

$$\tau_{13}^{\dagger}\tau_{51}^{\dagger}\tau_{16}\tau_{34}\vec{S}_{z} = \tau_{13}^{\dagger}(\tau_{51}^{\dagger}\tau_{16});\tau_{34}\vec{S}_{z} = \begin{pmatrix} 1 & 0 & 1 & 0 & 1 & 1 & 1 & 0 & 1 & 0 \\ 0 & 1 & 0 & 1 & 0 & 1 & 0 & 1 & 0 & 1 \\ 1 & 0 & 1 & 1 & 1 & 0 & 1 & 0 & 1 & 0 \\ 0 & 1 & 0 & 1 & 0 & 1 & 0 & 1 & 0 & 1 \\ 1 & 0 & 1 & 0 & 1 & 0 & 1 & 0 & 1 & 0 \\ 0 & 1 & 0 & 1 & 0 & 1 & 0 & 1 & 0 & 1 \\ 1 & 0 & 1 & 0 & 1 & 0 & 1 & 0 & 1 & 0 \\ 0 & 1 & 0 & 1 & 0 & 1 & 0 & 1 & 0 & 1 \\ 1 & 0 & 1 & 0 & 1 & 0 & 1 & 0 & 1 & 0 \\ 0 & 1 & 0 & 1 & 0 & 1 & 0 & 1 & 0 & 1 \end{pmatrix} \quad \ldots\ldots(16)$$

Our last result has brought out more interesting results. We can now see that operators $\tau_{ij}^{\dagger}\tau_{jk}$ are actually adding more particles into a particular spin state provided particles $a_{ij}$ and $a_{jk}$ were of the same spin state initially. However, no particles are added, to same spin state, whenever the selected particles are, initially, of different spin states. For a check on whether the trend is continuous, for higher order operation, we shall consider the following examples:

$$\tau_{54}^{\dagger}\left(\tau_{13}^{\dagger}\tau_{51}^{\dagger}\tau_{14}\tau_{32}\right)\tau_{41} = \begin{pmatrix} 0 & 1 & 0 & 0 & 0 & 1 & 0 & 1 & 0 & 1 \\ 1 & 0 & 1 & 0 & 1 & 0 & 1 & 0 & 1 & 0 \\ 0 & 0 & 0 & 1 & 0 & 1 & 0 & 1 & 0 & 1 \\ 0 & 0 & 1 & 0 & 1 & 0 & 1 & 0 & 1 & 0 \\ 0 & 1 & 0 & 1 & 0 & 1 & 0 & 1 & 0 & 1 \\ 1 & 0 & 1 & 0 & 1 & 0 & 1 & 0 & 1 & 0 \\ 0 & 1 & 0 & 1 & 0 & 1 & 0 & 1 & 0 & 1 \\ 1 & 0 & 1 & 0 & 1 & 0 & 1 & 0 & 1 & 0 \\ 0 & 1 & 0 & 1 & 0 & 1 & 0 & 1 & 0 & 1 \\ 1 & 0 & 1 & 0 & 1 & 0 & 1 & 0 & 1 & 0 \end{pmatrix} \quad \ldots\ldots(17)$$

$$\tau_{54}^{\dagger}\left(\tau_{13}^{\dagger}\tau_{51}^{\dagger}\tau_{16}\tau_{34}\right)\tau_{49} = \begin{pmatrix} 0 & 1 & 0 & 1 & 0 & 0 & 0 & 1 & 0 & 1 \\ 1 & 0 & 1 & 0 & 1 & 0 & 1 & 0 & 1 & 0 \\ 0 & 1 & 0 & 0 & 0 & 1 & 0 & 1 & 0 & 1 \\ 1 & 0 & 1 & 0 & 1 & 0 & 1 & 0 & 0 & 0 \\ 0 & 1 & 0 & 1 & 0 & 1 & 0 & 1 & 0 & 1 \\ 1 & 0 & 1 & 0 & 1 & 0 & 1 & 0 & 1 & 0 \\ 0 & 1 & 0 & 1 & 0 & 1 & 0 & 1 & 0 & 1 \\ 1 & 0 & 1 & 0 & 1 & 0 & 1 & 0 & 1 & 0 \\ 0 & 1 & 0 & 1 & 0 & 1 & 0 & 1 & 0 & 1 \\ 1 & 0 & 1 & 0 & 1 & 0 & 1 & 0 & 1 & 0 \end{pmatrix} \quad \ldots\ldots(18)$$

## 3. Result and Conclusion

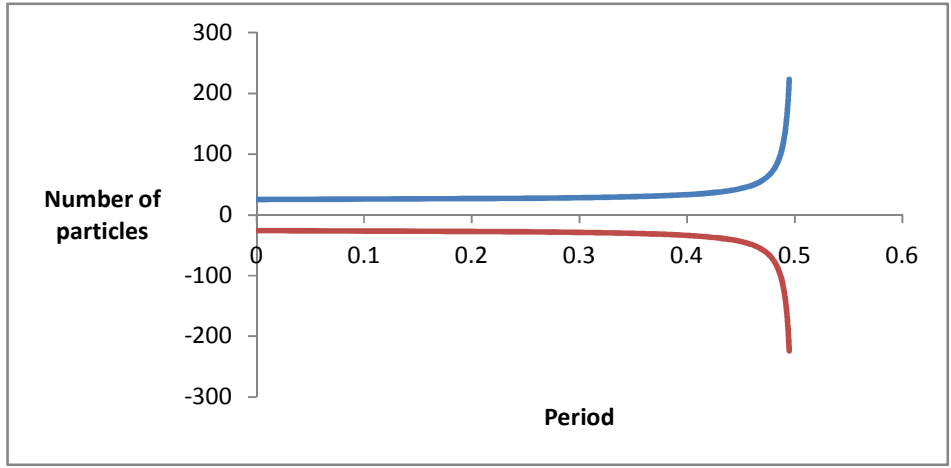

Fig 1: Steady growth in the population of same spin states. Blue-spin up, red-spin down. The negative values of the spin-down mean the opposite directions to spin up.

We can now formally write out a single equation to represent the combined operation of operators $\tau^{\dagger}_{ij}$ and $\tau_{ij}$ on our symmetric spin system. This of course, resembles the well-known occupation number

$$\hat{\eta} = \tau^{\dagger}_{ij\sigma} \tau_{jk\sigma} \ldots (19),$$ where $\sigma = \uparrow$ or $\downarrow$. Hence, the single spin operator for modeling the ferromagnetic alignment is the operator $\hat{\eta}$.

A many-body treatment for continuous higher-order promotion into same-spin state would be written as :

$$\hat{\eta} = \prod_{ij\uparrow \otimes jk\uparrow} \tau^{\dagger}_{ij\uparrow} \tau_{jk\uparrow} \ldots (20)$$

Going by the results obtained through equations (10-18) we have seen that the success of the model lies in the strong assumption that the acted upon particles must be, previously, of the same spin states. Hence, we can relate this steady growth in the population of same-spin state and thus represent the entire system of particles as a ferromagnet. We then show further with the graph above that the growth is steady and continuous for both spin states with time. The particles' alignment, in a specific direction, as stated by (20), steadily becomes uniform and synchronous with the alternation of the externally applied field. This description is just a wonderful way to model and explain ferromagnetism using spin symmetry.